\newcommand*{\RELEASE}{}  

\documentclass[conference]{IEEEtran}
\IEEEoverridecommandlockouts
\usepackage{cite}
\usepackage{amsmath,amssymb,amsfonts}
\usepackage{graphicx}
\usepackage{textcomp}
\usepackage{xcolor}
\def\BibTeX{{\rm B\kern-.05em{\sc i\kern-.025em b}\kern-.08em
    T\kern-.1667em\lower.7ex\hbox{E}\kern-.125emX}}

\usepackage{stfloats} 
\usepackage{algorithm}
\usepackage{algpseudocode}  

\usepackage{balance}

\usepackage{booktabs}  

\usepackage{tikz}
\usetikzlibrary{matrix,positioning}
\usetikzlibrary{decorations.markings}
\usetikzlibrary{shapes,arrows}
\usetikzlibrary{arrows.meta}

\graphicspath{{fig/}{../fig/}}

\usepackage[normalem]{ulem}  

\usepackage{colortbl}

\usepackage[utf8]{inputenc}
\usepackage{paralist}  
\usepackage{enumitem}  
\usepackage{xspace}  
\usepackage{url}

\definecolor{green(pigment)}{rgb}{0.0, 0.65, 0.31}

\ifdefined\RELEASE  
	\newcommand{\al}[1]{} 
	
	\newcommand{\del}[1]{}  
\else
	\newcommand{\al}[1]{\textcolor{green(pigment)}{[AL: #1]}} 
	
	\newcommand{\del}[1]{\textcolor{blue}{\sout{#1}}}  
\fi

\newcommand{\anm}{ANONYMOUS\xspace}
\newcommand{\kzalg}{SDA\xspace}  
\newcommand{\agghash}{AggHash\xspace}  
\newcommand{\errds}{40\%\xspace}  
\ifdefined\ANONYMOUS
	\newcommand{\urlTInfES}{\anm} 
	
	\newcommand{\urlStatix}{\anm}
\else
	\newcommand{\urlTInfES}{\url{https://github.com/eXascaleInfolab/TInfES}\xspace}
	
	\newcommand{\urlStatix}{\url{https://github.com/eXascaleInfolab/StaTIX}\xspace}
\fi
\newcommand{\ourtif}{StaTIX\xspace}  

\newcommand{\rlnmin}{5\xspace}  

\begin{document}


\title{\ourtif~---~Statistical Type Inference \\ on Linked Data
\thanks{
This project has received funding from the
\ifdefined\ANONYMOUS
\anm
\else
European Research Council (ERC) under the European Union’s Horizon 2020 research and innovation programme (grant agreement 683253/GraphInt) and in part by the Swiss National Science Foundation under grant number CRSII2 147609.
\fi
}
}


\author{
\ifdefined\ANONYMOUS
\\
\\
\anm\\
\\
\\
\else
\IEEEauthorblockN{Artem Lutov\IEEEauthorrefmark{1}, Soheil Roshankish\IEEEauthorrefmark{2}, Mourad Khayati\IEEEauthorrefmark{1} and Philippe Cudr{\'e}-Mauroux\IEEEauthorrefmark{1}}
\IEEEauthorblockA{\IEEEauthorrefmark{1}eXascale Infolab, University of Fribourg, Switzerland\\
Email: artem.lutov@unifr.ch, mourad.khayati@unifr.ch, pcm@unifr.ch}
\IEEEauthorblockA{\IEEEauthorrefmark{2}University of Bern, Switzerland\\
Email: soheil.roshankish@students.unibe.ch}
\fi
}

\maketitle

\begin{abstract}
Large knowledge bases typically contain data adhering to various schemas with incomplete and/or noisy type information. This seriously complicates further integration and post-processing efforts, as type information is crucial in correctly handling the data. In this paper, we introduce a novel statistical type inference method, called \ourtif, to effectively infer instance types in Linked Data sets in a fully unsupervised manner. Our inference technique leverages a new hierarchical clustering algorithm that is robust, highly effective, and scalable. We introduce a novel approach to reduce the processing complexity of the similarity matrix specifying the relations between various instances in the knowledge base. This approach speeds up the inference process while also improving the correctness of the inferred types due to the noise attenuation in the input data. We further optimize the clustering process by introducing a dedicated hash function that speeds up the inference process by orders of magnitude without negatively affecting its accuracy. Finally, we describe a new technique to identify representative clusters from the multi-scale output of our clustering algorithm to further improve the accuracy of the inferred types. We empirically evaluate our approach on several real-world datasets and compare it to the state of the art. Our results show that \ourtif is more efficient than existing methods (both in terms of speed and memory consumption) as well as more effective. \ourtif reduces the F1-score error of the predicted types by about \errds on average compared to the state of the art and improves the execution time by orders of magnitude.
\end{abstract}


\section{Introduction}
\label{sec:intro}

A significant fraction of the data available in knowledge bases today are stored as \emph{Linked Open Data (LOD)}. Large Linked Data projects such as the Linked Open Data Cloud~\footnote{\url{http://lod-cloud.net/}} or DBpedia~\cite{Aur07}
are often collaborative and contain data that has been extracted semi-automatically or that come from different sources. Hence, Linked Data often does not have a single maintainer or a strict unified schema for structuring the instances and as such typically includes noisy and/or incomplete data~\cite{Zave16}.
In particular, type information is often missing~\cite{Heik14}, which is particularly problematic as types are crucial for correctly handling many integration and post-processing tasks such as semantic search~\cite{Tono12}, federated query processing~\cite{Nol17}, linked data integration~\cite{Dut14}, or knowledge graph partitioning~\cite{Lhm17}. 

In this paper, we propose a novel method, called \ourtif~\footnote{\urlStatix} (for Statistical Type Inference), for automatically inferring instance types from Linked Data. Most methods in this context use supervised learning that leverage large, pre-labeled training sets (see the Related Work section). This can often turn out as a severe limitation, as such pre-labeled data are difficult and costly to acquire (or produce) for non-specialists or smaller entities, and as new labels are required for every new domain or dataset where type inference has to be applied. Our method instead is unsupervised and fully automated, and can be readily applied on any Linked Data source irrespective of its size or content.

\ourtif performs link-based statistical type inference leveraging a dedicated clustering algorithm that significantly improves type inference accuracy compared to the state of the art. Our link-based type inference technique takes as input weighted statistics from multiple attributes (properties) of each instance and avoids the propagation of errors from isolated erroneous axioms, similar to \cite{Heik13},
which allows it to operate on noisy data.
In particular, we propose a novel approach to simplify (\emph{reduce}) the processing complexity of the similarity matrix specifying the similarity between the instances. 
This reduction technique can speedup the clustering process by orders of magnitude, allowing to cluster larger datasets. Moreover, it can improve the overall accuracy of the results on noisy data. Also, we introduce a new optimization of the clusters formation process, using a dedicated hash function to speedup execution time by orders of magnitude without negatively affecting the resulting accuracy. Finally, we propose a novel technique to identify representative clusters from the resulting hierarchy, which further improves the accuracy of the type inference.

We perform an extensive empirical evaluation of our technique on real data and show that \ourtif significantly outperforms other unsupervised type inference approaches in terms of both effectiveness and efficiency. \ourtif reduces the accuracy error by about \errds on average comparing to other evaluated methods. In addition, \ourtif improves the execution speed by orders of magnitude and consumes less memory than the state of the art.

\section{Related Work}
\label{sec:relwork}


The classical way to perform type inference is the application of logical reasoning, e.g., via RDFS/OWL entailment regimes~\cite{Kaod08,Hors05}.
The resulting accuracy is highly dependent on the cleanliness and correctness of the statements in the knowledge base, though a number of works have attempted to reason on noisy semantic data~\cite{JiGH11}. Reasoning-based techniques are generally speaking considered as not suitable for cases where the knowledge base contains erroneous or conflicting statements~\cite{Heik13}. In addition, logical reasoning only allows to infer information from the facts that are present in the dataset; it is unsuited to infer types when most of the \emph{rdf:type} values are missing.

Several unsupervised type inference techniques have been proposed in the literature. In~\cite{Kenz15} the authors introduce a statistical method, which we refer to as \emph{\kzalg}, to compare the conformity of a dataset against its schema using statistical type inference. The proposed technique is based on the concept of probabilistic type profiles consisting of a set of properties $p$ and related probabilities $\alpha$ encoding the probability of an instance having $p$ as a property. In addition to the type profiles, a profile is assigned to each class in the schema to assess the completeness of the dataset and its conformity to the schema.
Paulheim et al.~\cite{Heik13} proposed a link-based classification technique, called \emph{SDType}, to infer missing types. SDType uses the statistical distribution of each link from an instance to assign types to instances. The statistical distribution is computed using a weighted voting approach, where a distribution of type votes is assigned to each link. The proposed technique outputs the confidence of each instance-type pair. SDType is implemented on top a relational database and achieves a quasilinear runtime complexity with the number of statements in the dataset.
It is important to outline that SDType requires some supporting database with ground-truth types (DBpedia is used by default), whose types are then assigned to the target dataset. Therefore, SDType aims solely at discovering  types that are present in the supporting dataset, even if those types have very little statistical significance in the target dataset, which conceptually differs from the semantics of the \kzalg results.

Both \kzalg and SDType are directly related to our present effort and are evaluated against our approach in the following.

A number of supervised techniques have been proposed in this context as well. Klieger et al~\cite{Klie16} proposed a supervised type inference technique called LHD~2.0 that extends the Linked Hypernyms Dataset (LHD) framework to extract types from DBpedia graphs. The proposed technique uses a statistical type inference (STI) technique to leverage the similarity between graphs by mapping classes appearing in one source knowledge graph, namely DBpedia, to another target knowledge graph, LHD. Together with the STI technique, the authors introduce an ontology-aware fusion approach based on hierarchical SVM to perform the assignment of instance types. LHD~2.0 combines a lexico-syntactic pattern analysis with supervised classification to assign the most probable types to the terms in the input text.
Zhang et al~\cite{Zhan17} introduced a data mining type prediction technique for Linked Data. The proposed technique is based on a text classification algorithm and boils down to a three-step procedure. First, a maximum entropy estimation is applied to find bags of words (BOW) in an RDF graph. Then, a weighted virtual document of type information (VDT) is computed. VDT consists of sub-BOW of words from URI of object $o_i$, sub-BOW of literals from $o_i$'s annotation properties and a sub-BOW of URIs of
$o_i$'s properties related to/from other objects. Each  sub-BOW is represented using word frequencies. This technique requires some \emph{a priori knowledge} on the number of target clusters and trains two classifiers to infer types.  
Bühmann et al.~\cite{Buhm16} introduced a system called DL-Learner to perform inductive learning on semantic web data. Their system provides an OWL-based machine learning tool to solve supervised learning tasks. It also supports knowledge engineers in constructing knowledge and learning about the data they created. A major component of this system is the induction process, which can be applied to infer types in a knowledge graph.
Melo et al.~\cite{Mel16} introduced a type prediction method called SLCN to tackle type incompleteness in Semantic Web knowledge bases with an ontology defining a type hierarchy. The authors formulate the type prediction problem as a hierarchical multi classification, where the class labels are types. The SLCN approach is based on a local classifier per node and performs feature selection, instance sampling, and class balancing. SLCN is applicable to large-scale RDF datasets with high-dimensional features.
%
The aforementioned supervised techniques were not designed to be applicable on cases where we do not expect any prior information on the schemas and instance types (e.g., when we do not know the number of types a priori and do not have any training set with corresponding labeled types), which is the focus of this paper (see Section~\ref{ssec:objective} for more detail).

\section{Method Overview}
\label{sec:method}

\subsection{Problem Statement}
\label{ssec:objective}

We consider Linked Data statements specified as RDF triples \emph{(s, p, o)}. Each triple is composed of three distinct components: a \emph{subject} (a URI or blank node), a \emph{predicate} (URI) and an \emph{object} (a URI, blank node or literal). 
In the following, we denote as an \emph{instance} $V_i$  
all triples sharing a given subject $i$. The predicates belonging to an instance correspond to attributes having a literal, a URI or a blank node as value. A special property defines the \emph{rdf:type} of $i$, indicating that $i$ is an instance of the class specified by the property value. Each instance may have multiple \emph{rdf:type} properties, i.e., may belong to multiple classes.
Our \emph{objective} is to infer missing \emph{rdf:type} values for all instances in the dataset \emph{G}, considering the following:
\begin{inparaenum}[\itshape a\upshape)]
\item the schema is incomplete; hence, both \emph{rdf:type} statements as well as class definitions may be missing;
\item classes can themselves be organized as to create a hierarchy (e.g., through \emph{rdfs:subClassOf} properties);
\item the dataset may be noisy (hence, \emph{G} may include incorrect statements).
\end{inparaenum}
In other words, we aim to induce types (\emph{rdf:type} values) for all instances, and classify each instance with respect to the discovered types for realistic scenarios where the data are both incomplete and noisy.

\subsection{Unsupervised Statistical Type Inference}  
\label{ssec:unsupervised}

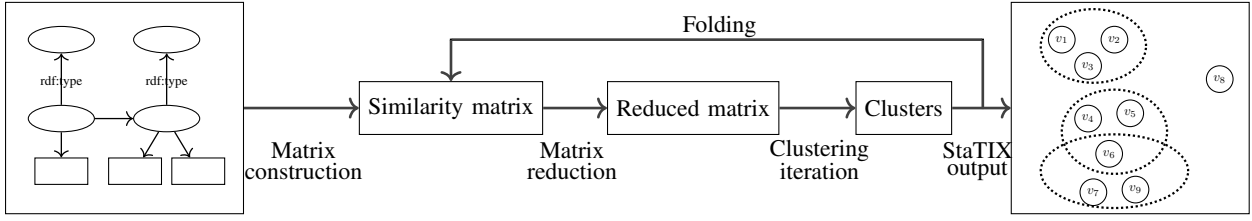
\begin{figure*}[t]\centering  
\begin{tikzpicture}
	[scale=.7,auto=left,every node/.style={circle,draw, scale=1}]



	\node[ellipse, minimum height=1em, minimum width=2.5em] (n1) at (-1,0) {};
			\node[ellipse, node distance=2.5cm, minimum height=1em, minimum width=2.5em] (n2) at (-1,-1.5)  {};
	\node[rectangle, minimum height=1em, minimum width=2em] (n3) at (-1,-2.5)  {};
	\node[scale =0.5, draw=none] (n12) at (-1,-0.8)  {rdf:type};

	\draw [<-, black, line width = 0.5 pt] (n1) edge node[draw =none]{} (n2);
	\draw [->, black, line width = 0.5 pt] (n2) edge node[left= -2pt, draw =none]{} (n3);

	\node[ellipse, minimum height=1em, minimum width=2.5em] (n5) at (1,0) {};
			\node[ellipse, node distance=2.5cm, minimum height=1em, minimum width=2.5em] (n6) at (1,-1.5)  {};
	\node[rectangle, minimum height=1em, minimum width=2em] (n7) at (0.4,-2.5)  {};
	\node[rectangle, minimum height=1em, minimum width=2em] (n8) at (1.6,-2.5)  {};
	\node[scale =0.5, draw=none] (n56) at (1,-0.8)  {rdf:type};

	\draw [<-, black, line width = 0.5 pt] (n5) edge node[draw =none]{} (n6);
	\draw [->, black, line width = 0.5 pt] (n6) edge node[left= -2pt, draw =none]{} (n7);
	\draw [->, black, line width = 0.5 pt] (n6) edge node[left= -2pt, draw =none]{} (n8);
	\draw [->, black, line width = 0.5 pt] (n2) edge node[left= -2pt, draw =none]{} (n6);

	\node[rectangle, minimum height=8em, minimum width=9em] (p1) at (0.2,-1.3)  {};

	\node[rectangle, minimum height=2em, minimum width=3em] (p2) at (6.4,-1.3)  {\small Similarity matrix};
	\draw[->, darkgray, line width = 1pt]  (p1) to (p2);
	\node[draw = none] (t11) at (3.6,-2.1)  {\small Matrix};
	\node[draw = none] (t12) at (3.6,-2.5)  {\small construction};

	\node[rectangle, minimum height=2em, minimum width=3em] (p3) at (11,-1.3)  {\small Reduced matrix};

	\draw[->, darkgray, line width = 1pt]  (p2) to (p3);
	\node[draw = none] (t21) at (8.7,-2.1)  {\small Matrix};
	\node[draw = none] (t22) at (8.7,-2.5)  {\small reduction};

	\node[rectangle, minimum height=2em, minimum width=3em] (p4) at (15,-1.3)  {\small Clusters};
	\draw[->, darkgray, line width = 1pt]  (p3) to (p4);
	\node[draw = none] (t3) at (13.4,-2.1)  {\small Clustering};
	\node[draw = none] (t3) at (13.4,-2.5)  {\small iteration};

	\node[rectangle, minimum height=8em, minimum width=9em] (p5) at (19.3,-1.3)  {};
	\node[scale =0.5] (v1) at (18,0) {$v_1$};
			\node[scale =0.5] (v2) at (19,0)  {$v_2$};
			\node[scale =0.5] (v3) at (18.5,-0.5)  {$v_3$};

			\node[scale =0.5] (v4) at (18.5,-1.5) {$v_4$};
			\node[scale =0.5] (v5) at (19.3,-1.4) {$v_5$};
			\node[scale =0.5] (v6) at (18.9,-2.17) {$v_6$};
			\node[scale =0.5] (v7) at (18.6,-2.9) {$v_7$};
			\node[scale =0.5] (v9) at (19.4,-2.85) {$v_9$};

			\node[scale =0.5] (v8) at (21,-0.75) {$v_8$};

	\node[draw = none] (t3) at (16.4,-2.1)  {\small \ourtif};
	\node[draw = none] (t3) at (16.4,-2.5)  {\small output};

			\node[rectangle, scale=0.03, fill=darkgray] (cl1) at (16.5,-1.3) {$v_8$};
			\node[rectangle, scale=0.03, fill=darkgray] (cl2) at (16.5,0) {$v_8$};
			\node[rectangle, scale=0.03, fill=darkgray] (cl3) at (6.4,0) {$v_8$};

	\draw[line width=0.3mm, densely dotted] (18.6,-0.12) ellipse (1cm and 0.7cm);
	\draw[line width=0.3mm, densely dotted] (19,-1.75) ellipse (1cm and 0.8cm);
	\draw[line width=0.3mm, densely dotted] (19,-2.5) ellipse (1.4cm and 0.7cm);

	\draw[->, darkgray, line width = 1pt]  (p4) to (p5);

	\draw[darkgray, line width = 1pt]  (cl1) to (cl2);
	\draw[darkgray, line width = 1pt]  (cl2) to (cl3);
	\draw[->, darkgray, line width = 1pt]  (cl3) to (p2);
	\node[draw = none] (t4) at (11.5,0.25)  {\small Folding
	};

\end{tikzpicture}
\vspace{-6pt}
\caption{Unsupervised Type Inference Process in \ourtif, where frames denote forms of the processing data and applied actions are displayed with arrows.}
\label{fig:typeinf}
\vspace{-6pt}
\end{figure*}

We now turn to a high-level description of our approach. 
We focus on a technique that is fully unsupervised and does not rely on any third-party knowledge base, and, therefore, can be readily applied to any Linked Data without any preparation or parameter tuning.
The fundamental assumption behind our approach is that the more properties the instances share, the more likely they 
have the same types.
Basically, we define the similarity between instances by matching their properties and then apply a dedicated clustering algorithm to infer the type clusters as shown in Fig.~\ref{fig:typeinf}.

Our type inference technique, \ourtif, takes a LOD dataset as input,
where 
some (or all) type information and class definitions are missing. From this input dataset, a \emph{Similarity Matrix} capturing the similarity among the instances is constructed, 
as explained below in Section~\ref{ssec:similarity}.
%
From there on, the type inference process is iterative, as the Similarity Matrix is reduced (see Section~\ref{ssec:lossred}) and clustered (see Section~\ref{ssec:clustering}) iteratively to infer 
clusters of types.

At the end of each iteration, weights for the resulting clusters as well as inter-cluster links are 
computed by aggregating the weight of the nodes in each cluster and the respective links (see \emph{folding} on Fig.~\ref{fig:typeinf}). The resulting (\emph{clustering}) graph forms a new Similarity Matrix and is used as a new input by the next iteration of the clustering algorithm.
The clustering process terminates as soon as an iteration does not produce any new cluster.
Clusters produced by this process form a hierarchy, where each level can be seen as representing the input dataset at a given level of granularity~\cite{Lcn09}. Clusters on the top (final) level of the hierarchy represent the inferred instance types. Inferring subtypes in addition to the the coarse-grained types requires considering the non-top level clusters, which is described in Section~\ref{ssec:clsmsc} and which improves the accuracy of the type inference.

The computational complexity of \ourtif varies from $O(m)$ on sparse clustering graphs\footnote{Note that a \emph{clustering} graph corresponds to the similarity matrix formed on the first iteration, which is typically much smaller than the input RDF graph} up to $O(m \sqrt{\frac{m}{n}})$ on dense noisy graphs, where $m$ is the number of links in the graph and $n$ is the number of nodes. The theoretical worst case corresponds to an extremely dense graph with weights yielding an excessive number of overlapping clusters and never occurs in practice thanks to our reduction technique (see below Section~\ref{ssec:lossred}).
The memory complexity of our approach depends on the same factors, and varies from linear on sparse graphs to
$O(m \cdot n)$ on the same worst case. We show that our technique is both space and time efficient in practice in Section~\ref{ssec:resconsum}.

\subsection{Similarity Matrix Construction}
\label{ssec:similarity}

The input of the clustering algorithm is a (clustering) graph, which formally can be represented by a similarity matrix. The matrix stores pairwise similarities between the instances in the input (RDF)
dataset. 
The similarities are computed like in \cite{Heik13} by applying a similarity function on vectors representing the set of properties attached to each instance. Both the property vectors and the similarity function are described below.

\subsubsection{Property Vectors}

Each instance in the input dataset is represented as a vector of its weighted properties. The weight $w_i$ of property $p_i$ expresses the importance of the property for the type inference and, intuitively, decreases for frequently occurring properties: $w_i = \frac{1}{\sqrt{freq_i}}$,
where $freq_i$ is the number of occurrences of $p_i$ in the dataset. We introduce the square root as the statistical distribution of links in real-world networks is typically heavy tailed~\cite{Barb16} and as we want to take into account a large number of properties (beyond the head of the distribution).
This weighting function 
yields better results than equal 
or frequency weighting in practice.

\subsubsection{Similarity Function}

Various functions, such as Cosine or Jaccard, can be used to evaluate the similarity between the property vectors. 
We use the 
cosine similarity 
as it is known to be highly effective~\cite{Levy15} and since it allows us to operate on weighted properties.  

\section{Type Inference} 
\label{sec:techniques}

We now turn to the core of our method. We describe below the three main steps of our type inference pipeline: Reduction of the Similarity Matrix (Section~\ref{ssec:lossred}), Clustering (Section~\ref{ssec:clustering}), and Cluster Identification at Multiple Scales (Section~\ref{ssec:clsmsc}).

\subsection{Weight-Preserving Reduction of the Similarity Matrix} 
\label{ssec:lossred}

Our novel similarity matrix reduction technique is applied before each clustering iteration to reduce the 
cost of the subsequent processing and to improve the accuracy of the results thanks to the link denoising.
The similarity matrix 
can be seen as an input graph for the clustering consisting of nodes (instances) and weighed links (pairwise similarities between the instances).
The number of links is in the worst case equal to the squared number of nodes---which only occurs when all pairs of instances share 
some 
property. 
However,
many links 
are insignificant in practice (as a given instance is typically related to a subset of the graph only) and as such can be omitted. 
Yet, carelessly removing lower-weight links can negatively impact the clustering process in two ways:
\begin{inparaenum}[\itshape a\upshape)]
\item a link connecting nodes \emph{A} and \emph{B} might be insignificant for node \emph{A} but significant for \emph{B}, and
\item the total weight of the graph should stay constant in order not to affect
the clustering of the remaining nodes (that are not adjacent to the nodes being reduced) when the \emph{global} optimization function (e.g., modularity~\cite{Nwm04u}) 
is applied.
\end{inparaenum}
The following reduction approach takes care of both cases.

Our reduction technique consists of two steps. First, we \emph{identify} insignificant links and then \emph{convert} them to weights of their respective nodes to retain the total weight of the input graph.
A link is considered as insignificant when it has no impact on the clustering of its incident nodes, which is defined by the optimization function of the clustering.
In the scope of this paper, we present a lightweight reduction approach, which is independent from the optimization function and operates on the link weights of the input graph directly.
Our approach 
is inspired by 
the empirical observation that clusters formed 
by picking the minimum-weight link a node
rarely maximize the optimization function.
Moreover, the higher the number of links connected to a node, the lower the probability that the minimum-weight link impacts cluster formation. Hence, two key values should be taken into account when reducing the graph:
\begin{inparaenum}[\itshape a\upshape)]
\item the minimal number of links a node should have to be qualified for the reduction without negatively affecting the clustering accuracy) and
\item the maximal number of links that can be reduced in a node to not (significantly) affect the clustering.
\end{inparaenum}

The minimal number of links a node should have to be eligible for the link reduction is defined formally considering the following aspects.
\begin{itemize}[leftmargin=*]
\item The performed reduction should not cause the formation of disconnected clusters (not linked to any node outside of the cluster). A cluster regroups together nodes with the most relevant relations, which roughly corresponds to the heaviest link weighs. Therefore, the non-reducible (head) links of the node should include the heaviest links with at least two distinct weights. 
\item A node link can be considered for the reduction only if its weight is insignificant, i.e. the weight is closer to zero than to the heaviest link weight of the node: $w_i < \frac{w_o}{2}$.
\end{itemize}
The reducible (tail) links of the node effectively consist of at least one link. Therefore, a node being eligible for link reduction includes at least two remaining links with distinct weights and one lightweight link being reduced, which strictly defines the hard threshold, 
$lsmin \ge 3$.
However, nodes having only three and even four links often do not satisfy our outlined restrictions.
So, empirically we select $lsmin = \rlnmin$ considering that the reduction for nodes having at most four links does not yield any speedup for our applied clustering algorithm. The actual number of reducible links is defined automatically as follows for the nodes having at least $lsmin$ links.

\begin{figure}[tbp]  
\centering
\includegraphics[scale=0.75]{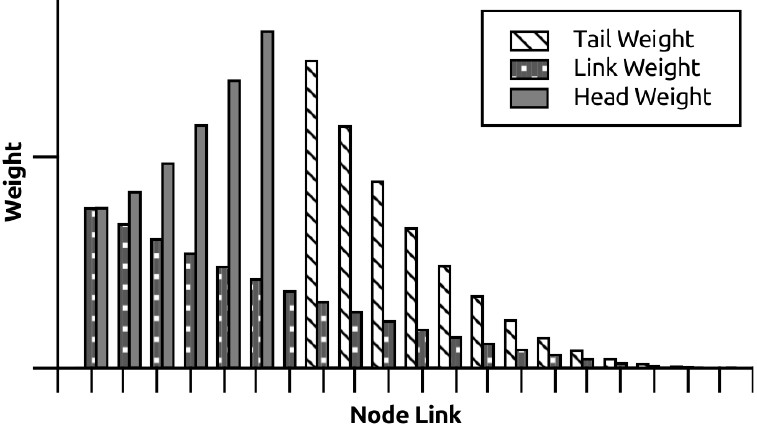}  
\caption{Node link weights and their accumulation from the tail to identify reduction candidates.}
\label{fig:ndlsrds}
\vspace{-6pt}
\end{figure}

\begin{algorithm}[H] 
\caption{Weight-preserving Similarity Matrix Reduction}
\label{alg:reduciton}
\begin{algorithmic}[1] 
\Procedure{reduceDensity}{$graph$}
\State $lsmin$ = \rlnmin 
\label{aln:lsmin}
\For{$node$ in $graph$ where \textnormal{count}(\textnormal{links}$(node)) \ge lsmin$ 
}
	\Comment{Identify reduction candidates}
	\State order(links(node))  \label{aln:lsorder}
	\State $els$ = end(links($node$)); $bls$ = begin(links($node$))  \label{aln:h2waccb}
	\State $ih$ = $bls$; $wh$ = weight($ih$); $wc$ = $wh$
	\For{$i$ in \textnormal{range}(2)}
		\While{++$ih \ne els$ and $wc \le \textnormal{weight}(ih)$}  
			\State $wh$ += rankedWeigh($ih$, $node$) \label{aln:hwacc0}
		\EndWhile
		\State $wc$ = weight($ih$)
	\EndFor
	\If{$ih$ == $els$}
		\State \textbf{continue}
	\EndIf  \label{aln:h2wacce}

	\State -\,-\,$ih$; $wtlmax$ = weight($bls$) / 2  \label{aln:lsaggb}
	\State -\,-\,($it$ = $els$); $wt$ = weight($it$) 
	\While{$it \ne ih$ and \textnormal{weight}$(it) < wtlmax$}
		\While{$wt < wh$ and \textnormal{weight}$(it) < wtlmax$ and $it \ne ih$}
			\State $wt$ += weight(-\,-\,$it$)
		\EndWhile
		\If{\textnormal{weight}$(it) < wtlmax$ and $it \ne ih$}
			\State $wh$ += rankedWeigh($ih$++, $node$) \label{aln:hwaccd}
		\EndIf
	\EndWhile
	\State $wc$ = weight($it$)
	\While{++$it \ne els$ and $wc \le \textnormal{weight}(it)$}  
	 \label{aln:dtiebr}
	\EndWhile
	\label{aln:lsagge}

	\For{$ln$ in \textnormal{range}($it$, $els$)}  \label{aln:lscvrb}
		\If{\textnormal{marked}($ln$)} \Comment{Convert insignif. links}
			\label{aln:rcdelbeg}
			\State addWeight(srcNode($ln$), weight($ln$) / 2)
			\State addWeight(dstNode($ln$), weight($ln$) / 2)
			\State remove($graph$, $ln$)
			\label{aln:rcdelend}
		\EndIf
		\State mark($ln$)  \label{aln:rcmark}
	\EndFor  \label{aln:lscvre}
\EndFor
\EndProcedure
\end{algorithmic}
\end{algorithm}
Our reduction technique is outlined in Algorithm~\ref{alg:reduciton} and illustrated in Fig.~\ref{fig:ndlsrds}.
First, we order the links of each node by descending weight on line~\ref{aln:lsorder} of Algorithm~\ref{alg:reduciton}. Then, we initialize the head links of the node by
accumulating all the links having the two heaviest weights (lines~\ref{aln:h2waccb}-\ref{aln:h2wacce}).
During the head link weight accumulation, the \texttt{rankedWeight} function (line~\ref{aln:hwacc0}) adopts an increasing ratio $rw_i \in (0, 1]$ starting from the second heaviest link ($i = 1$): $rw_i = \frac{2 i}{ndlsnum - 2}$,
where 
$1 \le i < \frac{ndlsnum}{2}$ 
and $ndlsnum \ge lsmin$ is the number of links in the node.
Afterwards, we iteratively aggregate the reduction candidates in the tail and the remained links in the head (lines~\ref{aln:lsaggb}-\ref{aln:lsagge}) till
\begin{inparaenum}[\itshape a\upshape)]
\item the tail has a lower weight than the head,
\item the tail can be expanded with links not assigned to the head and
\item the weight of each tail link is less than a half of the weight of the first head link $\frac{w_o}{2}$.  
\end{inparaenum}
Each iteration of the aggregation results in the addition of a single Head Weight bar and several Tail Weight bars until convergence as shown in Fig.~\ref{fig:ndlsrds}. The tail links being picked as reduction candidates are marked on line~\ref{aln:rcmark} for each node in the graph.
The links marked from both of their incident nodes are identified as insignificant and removed on lines~\ref{aln:lscvrb}-\ref{aln:lscvre} transferring their weights to their respective nodes to retain the total weight of the graph.

Through this reduction, the size of the respective
similarity matrix remains constant but the number of null values is increased, providing opportunities for more efficient storage and processing (only the non-null values are actually stored and processed by our system). Moreover, the reduction implicitly acts as a noise filtering step, which often improves the accuracy of the subsequent clustering as we show in Section~\ref{sec:results}.

\subsection{Clustering}  
\label{ssec:clustering}

We now turn to the unsupervised technique we use to infer 
clusters of instances sharing similar types. The problem definition that we consider imposes a number of requirements to the clustering algorithm. First, as an instance may have multiple types, we need an \emph{overlapping} (also called fuzzy or soft) clustering algorithm to allow instances to belong to several clusters (i.e., types). Second, as types may also form hierarchies, using a \emph{hierarchical} (or multi-scale) clustering technique would be desirable. Third, as we aim to infer types for any dataset without any manual labeling or tuning, the clustering algorithm should be \emph{parameter-free} (without any parameter to tune). Finally, as the input dataset might be noisy, the clustering algorithm should be \emph{robust}.
In addition to those criteria, the clustering technique should be efficient and scalable; both its time and space complexity should be lower than quadratic to be applicable to large datasets in practice. We developed a dedicated clustering algorithm to meet all those criteria. While a comprehensive presentation of our clustering algorithm itself is out of the scope of this paper, our implementation is available online as an open-source library 
and we provide a high-level description of the algorithm below.

We picked as the basis of our technique the Louvain clustering algorithm~\cite{Bld08}, which is a well-known community detection method to discover communities in large networks. Louvain is a \emph{greedy optimization} method that iteratively optimizes the \emph{modularity gain}~\cite{Nwm04u} of the resulting 
clusters.
\emph{Modularity gain} ($\Delta Q$) is shown in~\eqref{eq:dmod} and provides fast optimization of the modularity measure ($Q$) shown in~\eqref{eq:mod}. Formally, $\Delta Q$ captures the difference in modularity when merging two nodes \verb|#i| and \verb|#j| into the same cluster:
\begin{align}
\Delta Q_{i,j} &= \frac{1}{2w} \bigg( w_{i,j} - \frac{w_i w_j}{w} \bigg)
\label{eq:dmod}
\end{align}
\emph{Modularity} ($Q$)~\cite{Nwm04u} is a standard measure of clustering quality 
that is equal to the difference between the density of the links in the clusters and the expected density: 
\begin{align}  
Q &= \frac{1}{2w}\sum_{i,j}\left({w_{i,j} -\frac{w_i w_j}{2w}}\right)\delta(C_i, C_j)
\label{eq:mod}
\end{align}
where $w_{i,j}$ is the accumulated weight of the arcs (directed links) between nodes \verb|#i| and \verb|#j|, $w_i$ is the accumulated weight of all arcs of \verb|#i| (weight of each link is taken in each direction), $w$ is the accumulated weight of all edges (undirected links, half of the weight of all arcs) in the network, $C_i$ is the cluster to which \verb|#i| is assigned, and Kronecker delta $\delta(C_i, C_j)$ is a function, which is equal to $1$ when \verb|#i| and \verb|#j| belong to the same cluster (i.e., $C_i = C_j$), and $0$ otherwise.
Besides being extremely fast, modularity maximization under certain conditions is equivalent to the provably correct 
but computationally expensive methods of graph partitioning, spectral clustering and to the maximum likelihood method applied to the stochastic block model~\cite{Nwm13, Nmn16}.
The Louvain method is 
hierarchical, parameter-free, and efficient but it does not support overlapping clusters and is not robust. 
Hence, we modified the Louvain algorithm in the following ways.

\smallskip
\noindent{\bf{}Support for overlaps:} Overlaps occur when a node is shared by several clusters and has an equally good value with each of them in terms of the optimization function. To support overlaps, we transform the original graph to represent such cases explicitly in the network by \emph{decomposing} the node into (virtual) sub-nodes that can be processed independently in different clusters. This 
weight-preserving transformation does not influence the optimization function (as we ensure that the global modularity value stays constant), and allows to explicitly consider the fact that a single node can belong to several clusters.

\smallskip
\noindent{\bf{}Robustness:} Robustness implies stable results even if the input data are shuffled or are subject to minor perturbations (e.g., in the presence of noisy statements). Robust algorithms typically leverage some form of consensus (e.g. majority voting) to infer 
clusters by processing an input network multiple times and varying either the parameters of the clustering, the optimization function, or even the clustering algorithms in the meta-algorithm (e.g. OSLOM~\cite{Lnc12}). To avoid the high computational costs of such methods, we devised a new consensus approach on top of Louvain. The basic idea behind our new consensus approach is simple:
to cluster a pair of adjacent nodes together, we consider the (\emph{mutual}) maximal modularity gain from \emph{each} of the nodes instead of the maximal modularity gain from any of them. Thus, we apply a lightweight consensus approach, 
which yields robust (due to the consensus~\cite{Mnt03}) and fine-grained clusters at each level of the hierarchy.

\smallskip
\noindent{\bf{}Efficiency:}
A computationally heavy analysis has to be performed to decide whether a single or multiple overlapping clusters should be formed when a node has multiple clustering candidates (i.e., neighbors having the same mutual maximal value for the optimization function).
This analysis includes the identification of the minimal subset of clustering candidates (of the origin node) being also clustering candidates between each other. If such a subset exists, then a single solid cluster is formed comprising 
inter-node 
relations.
This analysis is the most computationally expensive step in our clustering process. However,
there is a frequent special case,
which can be identified and processed extremely efficiently. In semantic networks,
a node often has multiple clustering candidates and all of them are also clustering
candidates between each other. This case corresponds to a clear fine-grained subtype 
of some actual type (note that existing noise in the input data gets attenuated by the Similarity Matrix Reduction approach described above).

To identify such special cases with a reasonable efficiency (i.e. without having to order the clustering candidates and then comparing ordered sets) 
Bloom, Quotient Filter or any other (approximate) membership structures 
could be leveraged.
However, we propose a different, much more efficient approach, which is theoretically optimal and
yields a linear time complexity on the 
number of clustering candidates of the node. 
Our approach is a new \emph{history-independent} (i.e., independent of the input order) 
hash function, \agghash.
Generally, a history-independent hashing of a set can be achieved 
by applying any standard hash function to each item in the set with a subsequent 
commutative aggregate operator, which maintains the distribution of the hash values (e.g., \texttt{xor} for the uniform distribution).
However, the resulting hashing 
is required to minimize the number of collisions, since each collision causes omission of the respective fine-grained clusters and results in merging them into a single super cluster.
The latter requirement implies  
the application of some efficient cryptographic hash function, which is at least an order of magnitude slower than an effective non-cryptographic one.
Also, a non-cryptographic hash function typically results in less uniform distributions of the hash values, which boosts the number of collisions yielded by the aggregating \texttt{xor}.
To strike an ideal balance between accuracy and efficiency, we designed a dedicated hash function, \agghash.

\agghash is a cache-friendly, history-independent and 
aggregating hashing function for unordered sets. 
Being cache-friendly and history-independent, \agghash is an order of magnitude faster on CISC architectures than \texttt{xor} operations on generic non-cryptographic hash functions (e.g., MurmurHash from the C++ standard library).
Given a set $A$ of unique node ids $a_i$ of size $N_A$,
\agghash is defined as: 
\begin{multline}
\agghash(A) \rightarrow \{\{N_A, \sum_{i = 1}^{N_A} a_i, \sum_{i = 1}^{N_A} a_i^2\}\\
|\; i = \{1, \ldots, N_A\}\land a_i \in \mathbb{N}_0\}
\label{eq:agghash}
\end{multline}

\begin{algorithm}[bp]  
\caption{\agghash Hashing}
\label{alg:agghash}
\begin{algorithmic}[1]  
\Procedure{hashNode}{$node$}
\If{\textnormal{hashed}($node$)}
	\State \Return
\EndIf
\State $nid$ = corr(id($node$))  \label{aln:idcor} \Comment{Corrected node id}

\State $hash$ = \{1, $nid$, $nid \cdot nid$\}  \Comment{Init(num, sum, sum2)}  
\For{$cnode$ in \textnormal{clscands}($node$)}
	\State $hash.num$ += 1;\,
	$nid$ = corr(id($cnode$))  \label{aln:hashb}
	\State $hash.sum$ += $nid$;\,
	$hash.sum2$ += $nid \cdot nid$  \label{aln:hashe}
\EndFor
\State $node.hash$ = $hash$
\For{$cnode$ in \textnormal{clscands}($node$)}
	\State $cnode.hash$ = $hash$
\EndFor
\EndProcedure
\end{algorithmic}  
\end{algorithm}
The pseudo-code of \agghash is given in Algorithm~\ref{alg:agghash}.
In line~\ref{aln:idcor},  
\agghash performs a \emph{correction} of the input values to prevent collisions. This correction increments each input value by the square root of the maximal estimated input value (the largest possible node id in the input graph).
The intuition behind this transformation is the introduction of a lower bound for the partial values of the aggregating fields, since the probability of a collision drops quadratically with the increase of the smallest hashed value. 
The experiments we performed confirmed that this correction does not yield any collision on all datasets we processed. 
In lines~\ref{aln:hashb}-\ref{aln:hashe}, \agghash uses the addition operator (which is commutative), yielding an input nodes order-invariant result, i.e. performing a history-independent hashing for the clustering candidates (\texttt{clscands}) of the node.

\subsection{Representative Clusters Identification at Multiple Scales}
\label{ssec:clsmsc}

We propose a new technique to identify \emph{representative} clusters at multiple scales. We call a cluster representative if it is likely to represent an actual type, which happens only with a fraction of all clusters in the resulting hierarchy.
The representative clusters include
\begin{inparaenum}[\itshape a\upshape)]
\item the top level of the resulting hierarchy, which corresponds to coarse-grained types, and
\item some clusters on the lower levels (smaller scales), which correspond to fine-grained subtypes.
\end{inparaenum}
The intuition behind our selection technique is explained below.

\begin{figure}[tbp] 
\centering
\includegraphics[scale=1.1]{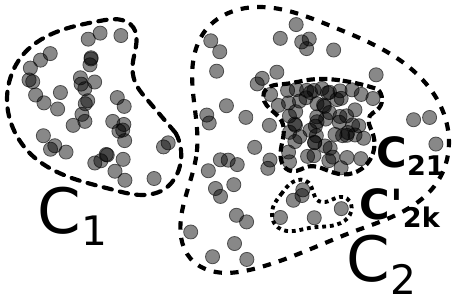}
\caption{$C_1$, $C_2$ and $C_{21}$ are representative clusters at various scales, $C'_{2k}$-like clusters are filtered out. 
}
\vspace{-6pt}
\label{fig:mscls}
\end{figure}
We illustrate our idea through an example. Fig.~\ref{fig:mscls} depicts a few clusters of nodes, where the similarity between the nodes is represented by their spatial closeness (density). Cluster $C_2$ consists of several sub-clusters ($C_{21}, C_{2k}, \ldots$) 
produced at lower levels of the hierarchy. 
Many sub-clusters, including $C'_{2k}$, have a density (i.e., strength of the pairwise instance similarity) lower than the average density of their super cluster $C_2$. Such a density variation is typical from real-world networks because of the heavy tailed distributions of node degrees and link weights~\cite{Barb16}, which result in a heavy-tailed distribution of cluster size and densities. The sub-clusters having a lower 
density than their super cluster typically do not represent groups that are statistically significantly different from their super cluster. Therefore, we select sub-clusters with a higher density of nodes only, which are likely to represent actual subtypes in the Linked Data.

In particular, we apply the following technique 
to retain only the most representative clusters, i.e., the inferred subtypes. Starting from the top level of the hierarchy, we evaluate the density of each cluster as its weight 
divided by its number of nodes. All sub-clusters having a density lower than  their direct super clusters or having a weight close to either
\begin{inparaenum}[\itshape a\upshape)]
\item 0 or
\item the weight of their super cluster, 
\end{inparaenum}
are filtered out as non-representative clusters. 
As a result, we end up with high-density clusters only, which have distinct statistical properties and are more likely to represent actual subtypes as we empirically show in the following section.

\section{Evaluation}
\label{sec:evals}

In this section, we first describe our experimental environment, including the evaluation metrics and datasets we used. Then, we present and discuss our results. The evaluation was performed on a Linux Ubuntu 16.04.3 LTS server with an Intel(R) Core(TM) i7-4770 CPU @ 3.40GHz (8 cores) and 32 GB RAM. Our evaluation framework (including all scripts and datasets) is open-source and available online~\footnote{\urlTInfES
}. 

\subsection{Metrics}
\label{ssec:metrics}

We initially assume that some (or all) type labels might be missing. We measure the accuracy of the resulting unlabeled types, which are represented by clusters of instances, using F1h (described below).
In case some type labels are available, we assign each label to the \emph{best-matching inferred cluster} using a weighted F1-score~\cite{Rjsb79} as matching criterion. 
Note that we assume that the labels might be missing and that the total number of types is unknown.

We measure the efficiency of the algorithms by measuring their runtime (in seconds) and their peak main memory consumption (in MB).

\subsubsection{F1h of Clusters}
\label{ssec:f1h}

The \emph{average F1-score (F1a)} is a common metric to measure the accuracy of clustering techniques~\cite{Yng13,Prat14}. F1a is defined as the average of the weighted F1-score~\cite{Rjsb79} of the \emph{best matching} ground-truth cluster to the inferred cluster and the F1-score of the \emph{best matching} inferred cluster to the ground-truth cluster. Formally, given the ground-truth clustering $C'$ consisting of clusters $c'_i \in C'$ and inferred clusters $c_i \in C$:
\begin{equation}
F1a(C',C) = \frac{1}{2}(F_{C',C} + F_{C,C'}),
\label{eq:f1a}
\end{equation}
where
\begin{multline}
F_{X,Y} = \frac{1}{|X|}\sum_{x_i \in X} F1(x_i, g(x_i, Y)),\\
g(x, Y) = \{\texttt{argmax}_y\; F1(x, y)\; |\; y \in Y\},
\label{eq:fxy}
\end{multline}
where $F1(x,y)$ is the F1-score for the respective clusters 
(subsets of the nodes $V$ of the input graph $G(V, E)$ represented by the similarity matrix).
This definition unfortunately yields \emph{non-indicative} values of $F1a \in [0, 0.5]$ for very large numbers of clusters, since the intentionally generated clusters representing all combinations of the nodes yield $F1a>0.5\, (F1_{C',C} = 1,\, F1_{C,C'} \rightarrow 0)$.
To address this issue, 
we use the harmonic mean instead, defining the \emph{harmonic F1-score (F1h)} as:
\begin{equation}
F1h(C',C) = \frac{2 F_{C',C} F_{C,C'}}{F_{C',C} + F_{C,C'}}.
\label{eq:f1h}
\end{equation}
$F1h \le F1a$ since the harmonic mean cannot be larger than the arithmetic mean.

\subsubsection{Weighted F1-score of Labeled Types (LF1)}
\label{ssec:lf1}

The F1-score together with Precision (P) and Recall (R) are a commonly used metrics for measuring the accuracy of labeled types. 
The weighted F1-score of the labeled types (LF1) represents the average F1-score of each labeled type ($g(l, C)$) weighted by the number of instances in the label ($|l|$):
\begin{equation}
LF1(C',C) = \frac{1}{|C'|} \sum_{l \in C'} |l| \frac{2 P_{l, g(l, C)} R_{l, g(l, C)}}{P_{l, g(l, C)} + R_{l, g(l, C)}}.
\label{eq:lf1}
\end{equation}
Note that each label can be assigned to several inferred types, that some inferred types might not have any assigned label, and that some types might have multiple labels. Thus, LF1 measures the accuracy of only the \emph{labeled types}, where F1h measures the accuracy of \emph{all resulting types}.

\subsection{Datasets}
\label{ssec:datasets}

\begin{table}[tbp]  
\centering
\caption{Evaluation Datasets.}
\label{tbl:data}
\begin{tabular}{@{}lrrrrc@{}}
\toprule
\multicolumn{1}{c}{\textbf{Dataset}} & \multicolumn{1}{c}{\textbf{Triples}} & \multicolumn{1}{c}{\textbf{Types}} & \multicolumn{1}{c}{\textbf{Nodes}} & \multicolumn{1}{c}{\textbf{Links}} & \textbf{Density} \\ \midrule
museum                      & 1418                         & 84                          & 178                       & 7143                      & 0.453   \\
soccerplayer                & 2654                         & 172                         & 272                       & 11008                     & 0.299   \\
country                     & 2273                         & 65                          & 453                       & 22176                     & 0.217   \\
politician                  & 3783                         & 200                         & 523                       & 32977                     & 0.242   \\
film                        & 6334                         & 5                           & 1303                      & 822557                    & 0.970   \\
mixen                       & 10128                        & 475                         & 1426                      & 244408                    & 0.241   \\
\rowcolor{gray!30} gendrgene                   & 5651                         & 7                           & 532                       & 140888                    & 0.997   \\
\rowcolor{gray!30} lsr                         & 56507                        & 11                          & 5767                      & 7621860                   & 0.458   \\
bauhist                     & 9022                         & 2                           & 861                       & 186460                    & 0.504   \\
schools                     & 15347                        & 3                           & 2256                      & 847320                    & 0.333   \\
histmunic                   & 119151                       & 14                          & 12132                     & 73380691                  & 0.997   \\ \bottomrule
\end{tabular}
\vspace{-6pt}
\end{table}
We consider three distinct categories of real-world Linked Open Data (LOD) datasets from various domains to evaluate and ensure a wide applicability of our unsupervised type inference.
It is worth outlining that a variety of data relations exist besides the distinct categories in the selected datasets. Each dataset contains multiple types, some datasets contain extremely diverse granularity of types while the granularity varies only slightly in others. Some of the types may contain a single instance even, which makes them statistically indistinguishable from noise. Instance types can be fully or partially overlapping with other types and some instances may not be attached to any type at all.
The first category of datasets are samples of DBpedia used for the \kzalg evaluation~\cite{Kenz15e}. 
This category is extended with \emph{mixen}, representing the union of the samples of the category datasets belonging to the English DBpedia. The second category are biomedical datasets~\footnote{\url{http://download.bio2rdf.org}} while the third category are open government datasets~\footnote{\url{https://opendata.swiss/en/dataset/}, \url{https://data.gov.uk/dataset/schools2}
}. Some statistics about each dataset are listed in Table~\ref{tbl:data}. The link density of the input graph is evaluated as the number of links (edges) divided by the maximal possible number of links ($nodes \cdot (nodes-1) / 2$).  

Note that not all those datasets define ground-truth types for all their instances. In case the ground-truth is missing for a given instance, we simply discard the instance when evaluating the F1 score.

\subsection{Results and Discussion}
\label{sec:results}

We compare both the effectiveness and the efficiency of our method against two state of the art unsupervised statistical type inference methods: \kzalg~\cite{Kenz15} and SDType~\cite{Heik13}. Additionally, we evaluate the impact of each proposed technique; \emph{\ourtif-rm-m-f} denotes the final results below, where \textit{rm} stands for similarity matrix reduction, \textit{m} for representative clusters identification, and \textit{f} for fast clusters formation using \agghash.
Since one of our main objectives is unsupervised type inference \emph{without any manual tuning}, the algorithms we evaluate are executed without any modification or parameters tuning. The latter is important for SDType, which is the only algorithm we evaluate requiring a supporting dataset. Moreover, SDType being tuned by default for DBpedia, it only considers incoming links for types discovery, while \ourtif and \kzalg consider all available links. However,
in order to not penalize SDType for types present in DBpedia but missing in the ground-truth, we discard such cases from the SDType results.

\subsubsection{Effectiveness}
\label{ssec:accuracy}

\begin{figure}[tbp]
\begin{minipage}[t]{0.48\textwidth}
\centering
\includegraphics[scale=0.65]{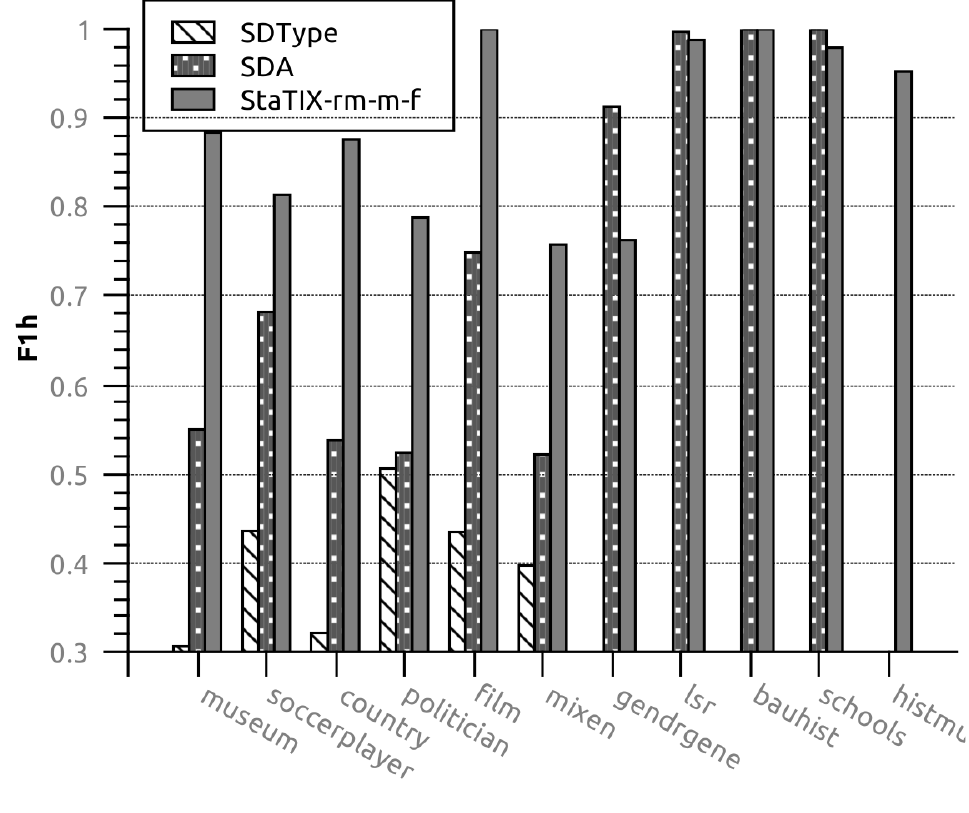}  
\vspace{-12pt}
\caption{Accuracy of the unsupervised statistical type inference algorithms by F1h measure.}
\label{fig:f1h}
\end{minipage}
\hfill
\vspace{12pt}
\begin{minipage}[t]{0.48\textwidth}
\centering
\includegraphics[scale=0.65]{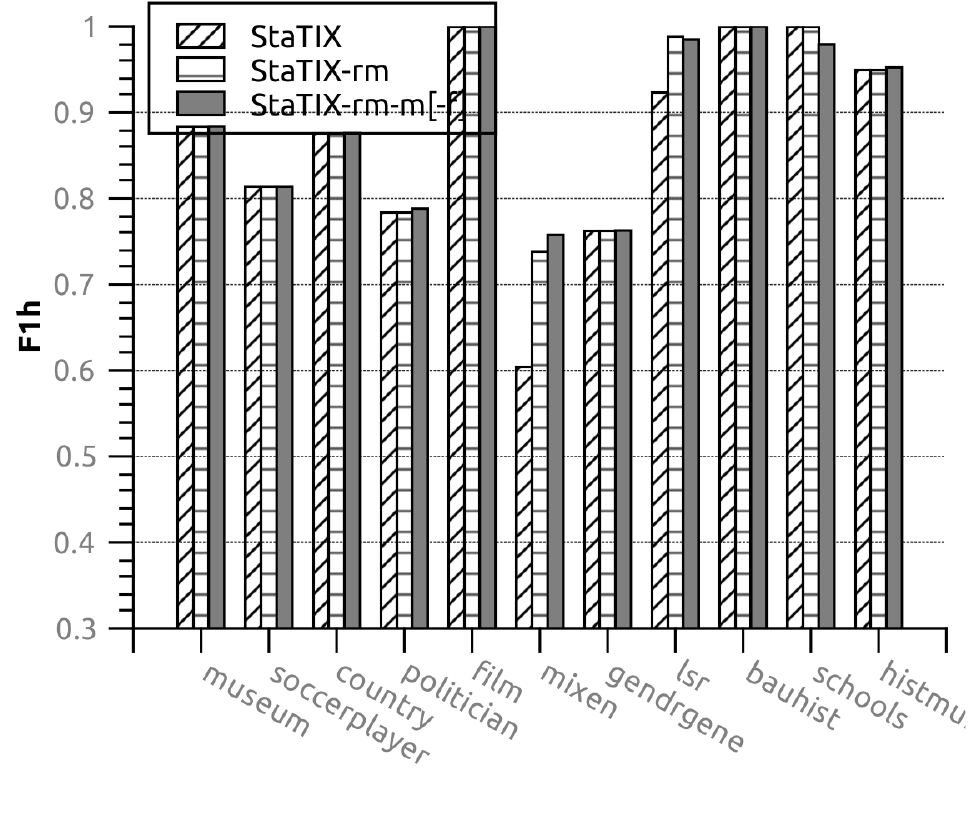}  
\vspace{-12pt}
\caption{Impact of the similarity matrix reduction technique on \ourtif accuracy by F1h measure.}
\label{fig:f1hsx}
\end{minipage}
\vspace{-6pt}
\end{figure}

Effectiveness results (in terms of \emph{F1h} score) are shown in Fig.~\ref{fig:f1h}. Our approach outperforms \kzalg reducing the F1h error by 37\% on average. 
On the last dataset, \emph{histmunic}, the results are absent for \kzalg as it
throws a Java heap space error
after 30 hours of evaluation.
\kzalg yields noticeably more accurate result than \ourtif does
 on a single dataset (\emph{gendrgene}) only, which has a ground-truth with heavily overlapping clusters of significantly varying sizes. Both \ourtif and \kzalg inferred 
the same number of types in this case, but \ourtif resolved overlaps more strictly by pruning some types that were correctly retained by \kzalg.
However, \kzalg has a relatively large variance in accuracy, which can be explained by the parameterized clustering algorithm (DBSCAN) used for the inference having default parameter values. In particular, \kzalg fails to detect large types with relatively weak relations between member instances, i.e. coarse-grained clusters of medium density, in half of the evaluated datasets. On the contrary, \ourtif yields a good accuracy with only small deviations from the ground-truth for all datasets independently of their size or density, thanks to our dedicated clustering algorithm.

The impact of our techniques on accuracy is shown in Fig.~\ref{fig:f1hsx}, where \ourtif-rm-m-f and \ourtif-rm-m are displayed with a single bar since the resulting clusters are exactly the same (\agghash does not affect the structure of the resulting clusters).
Our similarity matrix reduction approach significantly improves the accuracy on several datasets, which can be explained by the denoising effect, and 
in particular
by filtering out a number of noisy properties that caused the original clustering to get stuck on local optima.
The technique for representative clusters identification 
does not significantly impact the accuracy in terms of F1h, but improves the F1-score of the labeled types as shown in Table~\ref{tbl:labelsf1}. Most of the representative clusters correspond to fine-grained ground-truth labels, which otherwise are assigned to larger clusters and negatively impact recall. However, not all statistically representative clusters are present in the ground-truth, which penalizes F1h.
But even the statistically representative clusters that are absent in the ground-truth can be useful as they can help identify candidates for fine-grained types or outliers.
\begin{table*}[htbp]  
\centering
\caption{Accuracy of the labeled types by the weighted F1-score, 
the positive impact of the proposed techniques is outlined in bold.}
\label{tbl:labelsf1}
\begin{tabular}{@{}l|c|c|ccc|ccc|ccc@{}}
\toprule
\multicolumn{1}{c}{}
& \multicolumn{1}{c}{\ourtif} & \multicolumn{1}{c}{\ourtif-rm} & \multicolumn{3}{c}{\ourtif-rm-m[-f]} & \multicolumn{3}{c}{SDA} &  \multicolumn{3}{c}{SDType} \\
\cmidrule(lr){2-2} \cmidrule(lr){3-3} \cmidrule(lr){4-6} \cmidrule(lr){7-9} \cmidrule(lr){10-12}
\multicolumn{1}{l}{Dataset}
& \multicolumn{1}{c}{F1} & \multicolumn{1}{c}{F1} & \multicolumn{1}{c}{F1} & \multicolumn{1}{c}{P} & \multicolumn{1}{c}{R}  & F1 & P & \multicolumn{1}{c}{R} & F1 & P & R \\ \midrule
museum                       & 0.866                  & 0.866                  & 0.866                  & 1.000                 & 0.763 & 0.539 & 0.380 & 0.927 & 0.209 & 0.120 & 0.785 \\
soccerplayer                 & 0.789                  & 0.789                  & 0.789                  & 1.000                 & 0.652 & 0.695 & 0.574 & 0.882 & 0.447 & 0.339 & 0.657 \\
country                      & 0.840                  & 0.840                  & 0.840                  & 1.000                 & 0.725 & 0.632 & 0.478 & 0.930 & 0.249 & 0.155 & 0.634 \\
politician                   & 0.732                  & 0.732                  & \textbf{0.756}         & 0.982                 & 0.615 & 0.704 & 0.590 & 0.874 & 0.471 & 0.403 & 0.568 \\
film                         & 1.000                  & 1.000                  & 1.000                  & 1.000                 & 1.000 & 0.839 & 0.722 & 1.000 & 0.435 & 0.278 & 1.000 \\
mixen                        & 0.505                  & \textbf{0.723}                  & \textbf{0.751}         & 0.869                 & 0.662 & 0.559 & 0.412 & 0.873 & 0.378 & 0.360 & 0.398 \\
\rowcolor{gray!30} gendrgene                    & 0.806                  & 0.806                  & 0.806                  & 0.757                 & 0.861 & 0.889 & 0.987 & 0.809 &       &       &       \\
\rowcolor{gray!30} lsr                          & 0.912                  & \textbf{0.990}                  & 0.990                  & 1.000                 & 0.981 & 0.998 & 0.996 & 0.999 &       &       &       \\
bauhist                      & 1.000                  & 1.000                  & 1.000                  & 1.000                 & 1.000 & 1.000 & 1.000 & 1.000 &       &       &       \\
schools                      & 1.000                  & 1.000                  & 1.000                  & 1.000                 & 1.000 & 1.000 & 1.000 & 1.000 &       &       &       \\
histmunic                    & 0.950                  & 0.950                  & \textbf{0.958}         & 1.000                 & 0.920 &       &       &       &       &       &       \\ \bottomrule
\end{tabular}
\end{table*}

\subsubsection{Efficiency}
\label{ssec:resconsum}

\begin{figure}[bp]  
\centering
\includegraphics[scale=0.65]{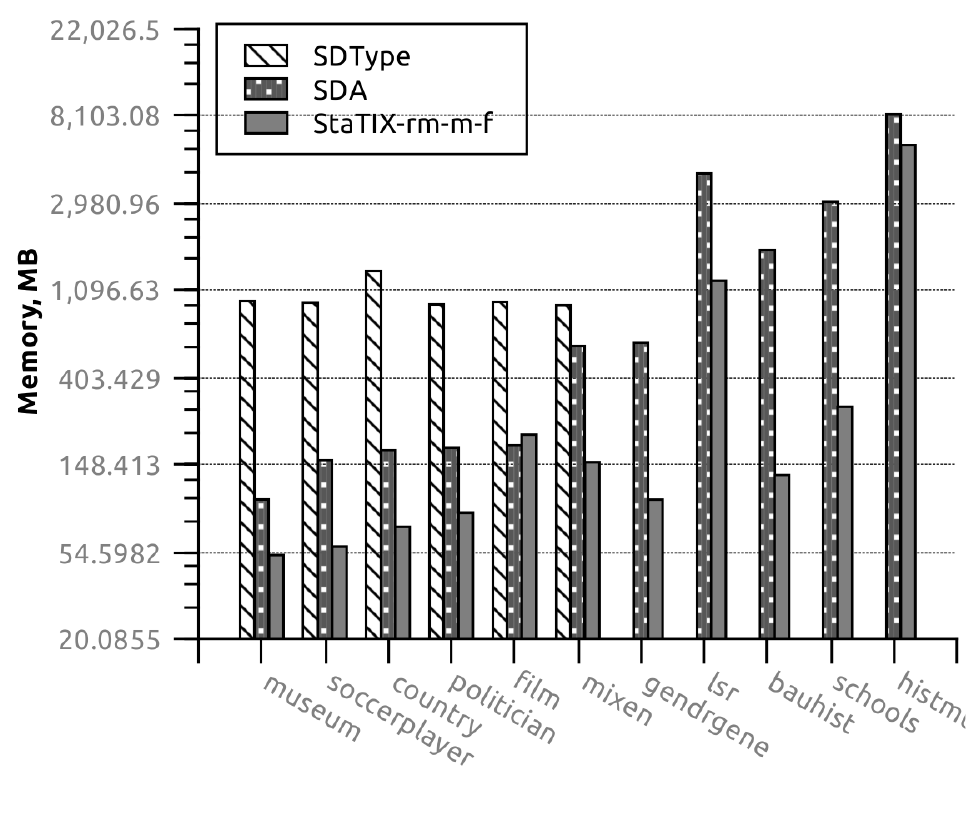}  
\vspace{-12pt}
\caption{Memory consumption of the unsupervised statistical type inference algorithms.}
\label{fig:memory}
\vspace{-6pt}
\end{figure}
In terms of space efficiency 
\ourtif consumes 1.5x-20x less memory than its competitors as shown in Fig.~\ref{fig:memory}. Memory consumption for SDType was not evaluated for the datasets where it produced empty results. Note that \kzalg throws a heap space error on the last dataset. \ourtif consumes slightly more memory than \kzalg on the \emph{film} dataset only, which is the smallest dataset having a density of links close to the theoretical maximum with a relatively small variation in terms of the link weights.
The memory overhead itself is caused by the deferred garbage collection in the JVM
of \ourtif resulting in the storage of two similarity matrices in memory 
when the matrix is streamed to the underlying native clustering library.
%
Our similarity matrix reduction technique speeds up subsequent clustering steps but does not affect the peak memory consumption since the matrix is generated and loaded in 
memory before being reduced.
To execute \ourtif on large LOD datasets, each column and row of the similarity matrix could be stored as a memory-mapped file on an SSD drive.

\begin{figure}[htbp]\centering
\centering
\begin{minipage}[t]{0.48\textwidth}
\centering
\includegraphics[scale=0.65]{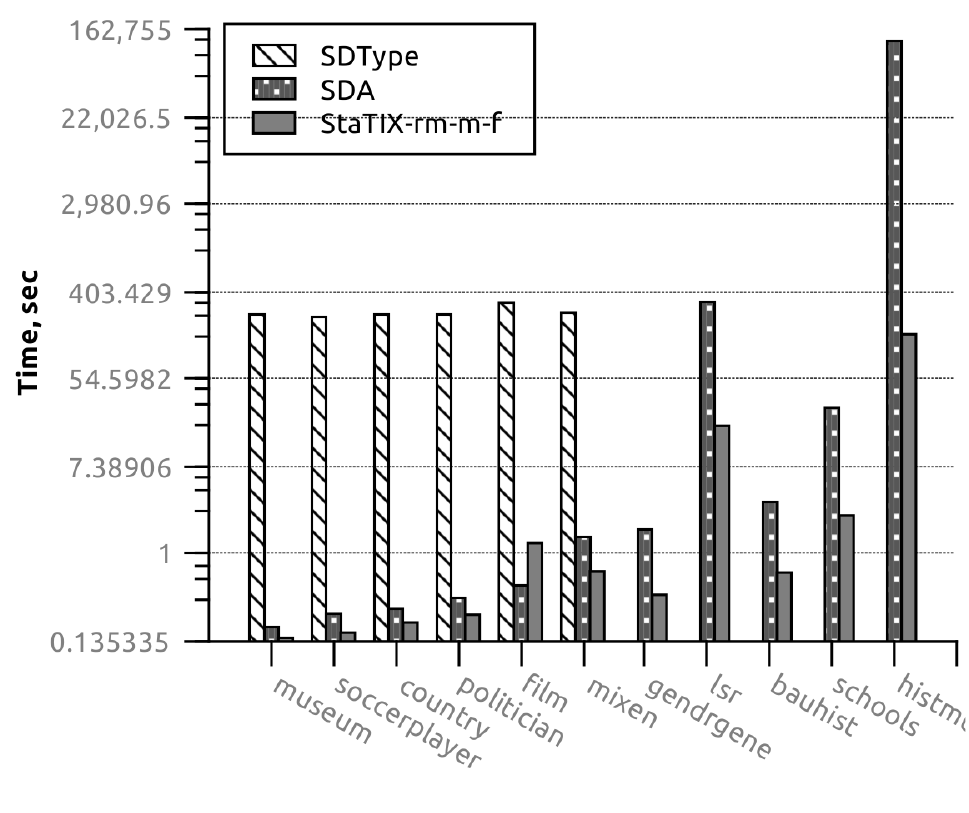}  
\vspace{-12pt}
\caption{Execution time of the unsupervised statistical type inference algorithms.}
\label{fig:exectime}
\end{minipage}
\hfill
\vspace{12pt}
\begin{minipage}[t]{0.48\textwidth}
\centering
\includegraphics[scale=0.65]{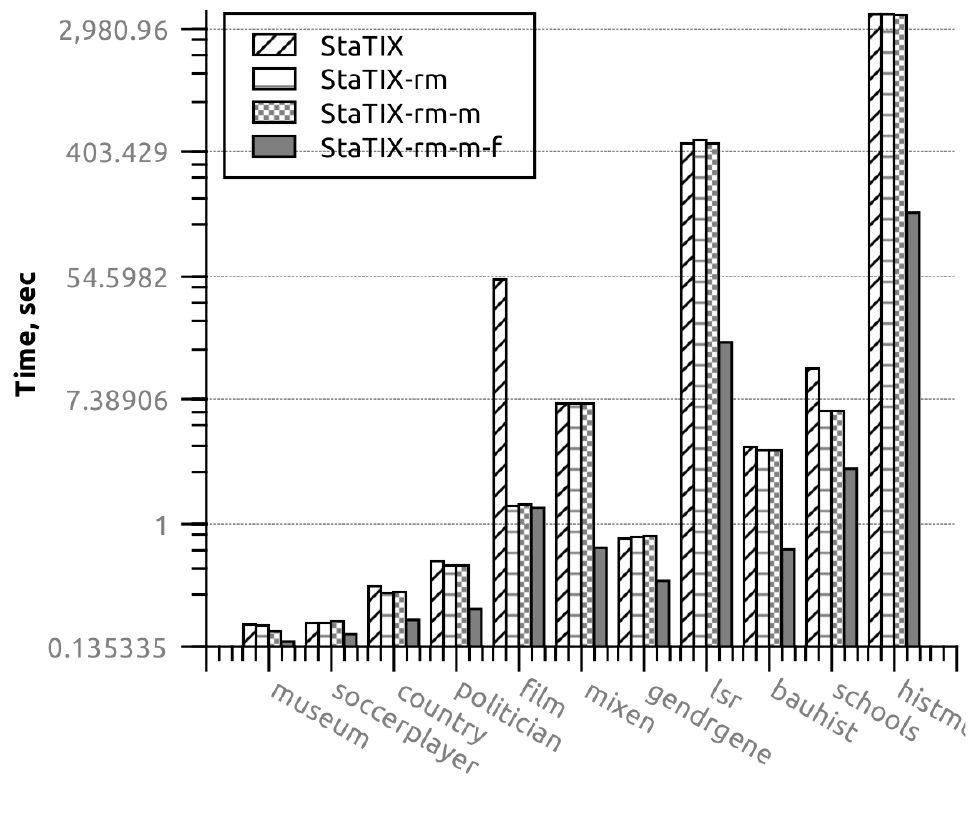}  
\vspace{-12pt}
\caption{Impact of the similarity matrix reduction technique on \ourtif execution time.}
\label{fig:etimesx}
\end{minipage}
\vspace{-4pt}
\end{figure}
The execution time of the algorithms 
is shown in Fig.~\ref{fig:exectime}, except for the cases where SDType produced empty results. 
\ourtif and SDType are implemented as single-threaded applications, whereas \kzalg takes advantage of multiple CPU cores. Nevertheless,
\ourtif performs type inference for the largest dataset (\emph{histmunic}) within 150 seconds, 
while \kzalg spends about 30 hours on up to 8 cores throwing an \emph{OutOfMemoryError} exception in the end.
In addition, we note that the type inference of \ourtif takes 35 seconds only, while most of the execution time is spent by \ourtif for I/O and RDF processing 
(consuming several GBs of memory also).
As shown in Fig.~\ref{fig:etimesx}, our similarity matrix reduction technique results in orders of magnitude speedups on several input datasets (e.g., 40x on the \emph{film} dataset). Fast clusters formation by \agghash speeds up  execution on all remaining datasets by a similar magnitude. Essentially, the execution time of \ourtif is bound by the I/O throughput and RDF conversion.

In conclusion, we improve over the state of the art by reducing the accuracy error by \errds on average and by reducing the execution time by up to three orders of magnitude 
on the evaluated datasets while requiring considerably less memory.

\section{Conclusions and Future Work}
\label{sec:conclusion}

In this paper, we introduced a new statistical type inference method, called \ourtif, to infer instance types in Linked Data in a fully automatic manner without requiring any prior knowledge about the dataset. Our method is based on a new clustering technique, which infers (overlapping) types in a robust and efficient manner. As part of our method, we also presented novel techniques to
\begin{inparaenum}[\itshape a\upshape)]
\item reduce the similarity matrix representing relationships between the instances,
\item speed up the clusters formation using a dedicated, history-independent hash, \agghash, and
\item identify representative clusters at multiple scales.
\end{inparaenum}
We empirically compared our approach on a number of different datasets and showed that it is at the same time considerably more efficient and orders of magnitude more effective than state-of-the-art techniques.

In the future, we plan to extend \ourtif with additional semantic analysis leveraging both logical reasoning and embedding techniques to better grasp the differences and relationships between various instances. We also plan to add support for automatically borrowing type labels from third-party knowledge bases whenever available.
In terms of implementation-specific aspects, we plan to parallelize our algorithm to take advantage of modern multi-core CPU architectures.

\bibliographystyle{IEEEtran}\balance
\bibliography{IEEEabrv,./statix}

\end{document}